\documentclass[12pt]{article} 
\usepackage{epsfig}
\usepackage{changebar} 
\usepackage{amssymb} 
\usepackage{amsmath}
\usepackage{amsbsy} 
\usepackage{subeqnarray}

\allowdisplaybreaks[4]

\def\intq#1{\int \frac{d^4#1}{(2\pi)^4}}
\def\intt#1{\int \frac{d^3#1}{(2\pi)^3}}

\def\intn#1#21{\int \frac{d^#2#1}{(2\pi)^#2}}

\def\Tr{{\rm Tr}}

\def\slash#1{\rlap{\hskip1pt /}#1}

\def\p{{|{\bf p}|}}
\def\q{{|{\bf q}|}}

\begin{document}

\title{On the stability of Quantum Hadro-Dynamics}

\author{M. B. Barbaro${}^\dagger$,  R. Cenni${}^\ddagger$, 
M. R. Quaglia${}^\ddagger$,\\
  ${}^\ddagger$ Istituto Nazionale di Fisica Nucleare -- Sez. di Genova\\
  Dipartimento di Fisica dell'Universit\`a di Genova\\
  ${}^\dagger$ Istituto Nazionale di Fisica Nucleare -- Sez. di Torino\\
  Dipartimento di Fisica Teorica dell'Universit\`a di Torino}

\date{ }
\maketitle

\begin{abstract}
We explore the possible occurrence of $\sigma$-$\omega$ condensation in the
Quantum Hadro-Dynamics (QHD), namely the Serot and Walecka model, 
finding that at the mean field level it corresponds
to a critical value of the coupling constant $g_\sigma=8.828$ and density
$k_F=207.2$ MeV/c, significantly below the standard value of QHD.
\end{abstract}

\section{Introduction}
\label{sec:1}

The Quantum Hadro-Dynamics (QHD) is a largely used model to describe 
nuclear systems within a coherent and covariant frame. Its
parameters ($\sigma$ and $\omega$ masses and coupling constants)
are tuned to reproduce the energy and density of the nuclear matter at
the mean field level, where  exotic or non-trivial phenomena are
intrinsically forbidden.

In going beyond the mean field, however,
one could question above the stability 
of the nuclear matter against the occurrence of a $\sigma$-$\omega$
condensate.

We wish to establish in this paper the limits the stability imposes
on the model parameters.

\section{General formalism}
\label{sec:2}

To shortly remember the QHD scheme \cite{SeWa-86}, 
we introduce the lagrangian density
\begin{equation}
  \label{eq:x1}
  {\cal L}={\cal L}_N+{\cal L}_\sigma+{\cal L}_\omega+{\cal L}_I
\end{equation}
with
\begin{eqnarray}
  \label{eq:x2}
  {\cal L}_N&=&\overline{\psi}(i\not\partial-m)\psi\\
  {\cal L}_\sigma&=&\frac{1}{2}(\partial^\mu\sigma)^2-\frac{1}{2}m_\sigma^2
  \sigma^2\\
  {\cal L}_\omega&=&-\frac{1}{4}F_{\mu\nu}F^{\mu\nu}+\frac{1}{2}m_\omega^2
  \omega_\mu\omega^\mu\\
  {\cal L}_I&=&g_\sigma\overline{\psi}\sigma\psi-g_\omega\overline{\psi}
  \gamma_\mu\psi\omega^\mu-\frac{1}{4!}a_4\sigma^4
\end{eqnarray}
and 
\begin{equation}
  \label{eq:x3}
  F^{\mu\nu}=\partial^\mu\omega^\nu-\partial^\nu\omega^\mu~.
\end{equation}

We get rid of the non-vanishing $\sigma$ and $\omega^0$ fields in the
vacuum
by means of the shifts
$\sigma=\sigma'-\bar\sigma$, $\omega^\mu\to{\omega^\mu}'-\bar\omega^\mu$,
the mean fields being determined 
by solving the classical equations of motion:
\begin{subeqnarray}
  \label{eq:x6}
  m_\sigma^2\bar\sigma+\frac{a_4}{3!}\bar\sigma^3
  &=&g_\sigma<\overline{\psi}\psi>\\
  m_\omega^2\bar\omega^0&=&g_\omega<\overline{\psi}\gamma_0\psi>\\
  {\bar\omega}^i&=&0
\end{subeqnarray}
Their solutions  read
\begin{equation}
  \label{eq:062}
  \bar\omega^\mu=\frac{g_\omega\rho}{m_\omega^2}\delta^{\mu 0}
\end{equation}
and
\begin{equation}
  \label{eq:x102}
  \bar\sigma
  =-\frac{2m_\sigma^2}{\sqrt{a_4}R_\sigma}+\frac{R_\sigma}{\sqrt{a_4}}~,
\end{equation}
where $\rho$ is the usual nuclear density
\begin{equation}
  \rho=\frac{2k_F^3}{3\pi^2}
\end{equation}
and
\begin{equation}
  \label{eq:x103}
  R_\sigma=\sqrt[3]{3\sqrt{a_4}g_\sigma\rho_\sigma+\sqrt{8m_\sigma^6+
      9a_4g_\sigma^2\rho_\sigma^2}}
\end{equation}
with
\begin{equation}
  \label{eq:064}
  \rho_\sigma=-i\Tr\intq{k}S_H(k)~.
\end{equation}
In the latter definition $S_H$ denotes a fermion propagator analogous
to the nucleon propagator in the medium
\begin{equation}
  \label{eq:019}
  S^0(k)=\frac{\slash k+m}{2E_k}
  \left\{\frac{\theta(k-k_F)}{k_0-E_k+i\eta}+
    \frac{\theta(k_F-k)}{k_0-E_k-i\eta}
      -\frac{1}{k_0+E_k-i\eta}\right\}
\end{equation}
but with the nucleon mass replaced by an effective one, 
defined by 
\begin{equation}
  \label{eq:Y001}
  m^*=m-\bar\sigma~.
\end{equation}

Eqs. \eqref{eq:x102} and
\eqref{eq:x103} need  to be solved self-consistently in $\bar\sigma$.
Its knowledge immediately provides the effective nucleon mass.
It is interesting to note that a self-consistent calculation realizes 
the prediction of Lee and Wick \cite{LeWi-74} that for large $k_F$ the nucleon
mass is vanishing.

\section{The Random Phase Approximation (RPA)}
\label{sec:5}

It has recently been observed\cite{Du-03} that 
the presence of a nuclear medium couples the $\sigma$ to the $0^{\rm th}$
component of the $\omega$, so that the $\sigma$ and $\omega$ propagation
in the medium will be described by two coupled Dyson equations.
We can write them  (in momentum space)
as a single equation in a 5-dimensional space in the form
\begin{equation}
  \label{eq:065}
  D^{(5)}=D_0^{(5)}+D_0^{(5)}\Pi^{*(5)}D^{(5)}
\end{equation}
where $D_0^{(5)}$, $D^{(5)}$ and $\Pi^{*(5)}$ are $5\times5$ matrices
having the structure
\begin{equation}
  \label{eq:066}
  D_0^{(5)}=
  \begin{pmatrix}
    D_{0\sigma}&0\\
    0&D^{\mu\nu}_{0\omega}
  \end{pmatrix}
  ~~~~
  D^{(5)}=
  \begin{pmatrix}
    D_{\sigma\sigma}&D_{\sigma\omega}^\mu\\
    D_{\omega\sigma}^\nu&D^{\mu\nu}_{\omega\omega}
  \end{pmatrix}
  ~~~~
  \Pi^{*(5)}=
  \begin{pmatrix}
    \Pi_{\sigma\sigma}^*&\Pi^{*\mu}_{\sigma\omega}\\
    \Pi^{*\nu}_{\omega\sigma}&\Pi^{*\mu\nu}_{\omega\omega}
  \end{pmatrix}
  ~.
\end{equation}
The free meson propagators are defined as
\begin{eqnarray}
  \label{eq:067}
  D_{0\sigma}(q)&=&g_\sigma\frac{1}{q^2-m_\sigma^2+i\epsilon}
  g_\sigma~,\\
  \label{eq:Z067}
  D^{\mu\nu}_{0\omega}(q)&=&-g_\omega\frac{g^{\mu\nu}-
    \dfrac{q^\mu q^\nu}{q^2}}
  {q^2-m^2_\omega+i\epsilon}g_\omega\equiv 
  -\left(g^{\mu\nu}-\dfrac{q^\mu q^\nu}{q^2}\right)D_{0\omega}(q)~.
\end{eqnarray}
The polarisation propagators are constrained by current conservation. Thus
they must take the form
\begin{equation}
  \label{eq:A073}
  \Pi_{\sigma\omega}^\mu(q)=\Pi_{\sigma\omega}^0(q){\mathfrak N}^\mu
  \equiv\Pi^V(q){\mathfrak N}^\mu~,
\end{equation}
with
\begin{equation}
  \label{eq:A074}
  {\mathfrak N}^\mu=\left(1,\frac{q^0q^i}{{\bf q}^2}\right)~,
\end{equation}
and
\begin{equation}
  \label{eq:A076}
  \Pi_{\omega\omega}^{\mu\nu}(q)=\left(
  \begin{tabular}{c|c}
    $\Pi^{L}$&$\frac{q_0q_i}{\q^2}\Pi^{L}$\\
    \hline
    $\frac{q_0q_j}{\q^2}\Pi^{L}$
    &$\frac{q_0^2}{\q^2}\Pi^{L}\frac{q_iq_j}{\q^2}
    +\frac{1}{2}\Pi^{T}\left(\delta_{ij}-\frac{q_iq_j}{\q^2} \right)
    $
  \end{tabular}\right)~,
\end{equation}
with 
\begin{equation}
  \label{eq:Z001}
  \Pi^L=\Pi_{\omega\omega}^{00}(q)
\end{equation}
and
\begin{equation}
  \label{eq:Z002}
  \Pi^T=\left(\delta^{ij}-\frac{q^iq^j}{\q^2}\right)\Pi_{\omega\omega}^{ij}(q)
  ~.
\end{equation}
The quantity  $\Pi_{\sigma\sigma}^*(q)$ has no tensor
structure. We shall define
\begin{equation}
  \label{eq:Z003}
  \Pi^S= \Pi_{\sigma\sigma}^*
\end{equation}
in order to harmonise the notations.

Further, since $q_\mu D^\mu_{\sigma\omega}=q_\mu D^\mu_{\omega\sigma}=0$
and $q_\mu D^{\mu\nu}_{\omega\omega}=0$, as Dyson equation implies, we must
also have
\begin{equation}
  \label{eq:A079}
  D^\mu_{\sigma\omega}=D^\mu_{\omega\sigma}=D^V{\mathfrak N}^\mu
\end{equation}
and
\begin{equation}
  \label{eq:A080}
  D^{\mu\nu}_{\omega\omega}=\left(
  \begin{tabular}{c|c}
    $D^{L}$&$\frac{q_0q_i}{\q^2}D^{L}$\\
    \hline
    $\frac{q_0q_j}{\q^2}D^{L}$
    &$\frac{q_0^2}{\q^2}D^{L}\frac{q_iq_j}{\q^2}
    +\frac{1}{2}D^{T}\left(\delta_{ij}-\frac{q_iq_j}{\q^2} \right)
    $
  \end{tabular}\right)~.
\end{equation}

The polarisation propagators needs to be approximated in some way.
It has been proved in \cite{AlCeMoSa-87} 
that at the level of  mean field
in a bosonic space they must be replaced with their zero-order
approximation. Thus we replace $\Pi^*$ with $\Pi^0$, with components
(traces refer both to
spin and isospin)
\begin{align}
  \label{eq:A072}
  \Pi_{0\sigma\sigma}(q)&=-i\Tr\intq{p}S_H(p)S_H(p+q)~,\\
  \Pi_{0\sigma\omega}^\mu(q)&=-i\Tr\intq{p}S_H(p)S_H(p+q)\gamma^\mu\\
\intertext{and}
  \Pi_{0\omega\omega}^{\mu\nu}(q)&=-i\Tr\intq{p}S_H(p)\gamma^\mu S_H(p+q)
  \gamma^\nu~.
\end{align}

The $0^{\rm th}$ order propagators $\Pi_0^{S(VLT)}$ 
have been extensively studied in ref. \cite{BaCeQu-05},
where their analytical representation is given. 
Here we observe that they can be written as
\begin{eqnarray}
  \label{eq:155}
  \Pi_0^S&=&4(4m^2-q^2)\Pi^0+8{\mathfrak T}\\
  \Pi_0^V&=&16m{\mathfrak Q}^V\\
  \Pi_0^L&=&16{\mathfrak Q}^L-4\q^2\Pi^0+8\frac{\q^2}{q^2}{\mathfrak T}\\
  \Pi_0^T&=&16{\mathfrak Q}^T+8q^2\Pi^0-16{\mathfrak T}~,
\end{eqnarray}
where
\begin{eqnarray}
  \label{eq:A092}
  \Pi^0&=&\intt{p}\frac{1}{2E_\mathbf{p}}
  \frac{\theta(k_F-p)}{(\tilde q_0+E_\mathbf{p})^2-E_{\mathbf{p}
      +\mathbf{q}}^2}\Biggm|_{p_0=E_\mathbf{p}}+\left(
q_0\longleftrightarrow -q_0\right)~,\\
  \label{eq:A292}
  {\mathfrak Q}^V&=&\intt{p}\frac{t^0}{2E_\mathbf{p}}
  \frac{\theta(k_F-p)}{(\tilde q_0+E_\mathbf{p})^2-E_{\mathbf{p}
      +\mathbf{q}}^2}\Biggm|_{p_0=E_\mathbf{p}}+
\left(q_0\longleftrightarrow -q_0\right)~,\\
  \label{eq:A392}
  {\mathfrak Q}^L&=&\intt{p}\frac{(t^0)^2}{2E_\mathbf{p}}
  \frac{\theta(k_F-p)}{(\tilde q_0+E_\mathbf{p})^2-E_{\mathbf{p}
      +\mathbf{q}}^2}\Biggm|_{p_0=E_\mathbf{p}}+
\left(q_0\longleftrightarrow -q_0\right)~,\\
    \label{eq:A492}
 {\mathfrak Q}^T &=&\intt{p}\frac{\p^2\q^2-({\bf }p\cdot{\bf q})^2}
 {2E_\mathbf{p}\q^2}
  \frac{\theta(k_F-p)}{(\tilde q_0+E_\mathbf{p})^2-E_{\mathbf{p}
      +\mathbf{q}}^2}\Biggm|_{p_0=E_\mathbf{p}}+
\left(q_0\longleftrightarrow -q_0\right)~,
  \nonumber
  \\
  ~\\
  {\mathfrak T}&=&\intt{p}\frac{\theta(k_F-p)}{2E_\mathbf{p}}~.
\end{eqnarray}
In the above we have introduced the transverse vector
\begin{equation}
  \label{eq:A151}
  t^\mu=p^\mu-\frac{p\cdot q}{q^2}q^\mu~,
\end{equation}
while 
\begin{equation}
  \label{eq:A152}
  \tilde q_0=q_0+i\eta~ {\rm sign}(q_0)
\end{equation}
accounts for the right analytical determination near the cuts.

We can now
 solve the Dyson equation \eqref{eq:065}, finding
\begin{align}
  \label{eq:A081}
  D_{\sigma\sigma}&=\widetilde D_\sigma
  \dfrac{1}{1-\Pi_0^V\widetilde D_\omega\Pi_0^V
    \widetilde D_\sigma}\\
  \label{eq:Z010}
  D^L&=\left(\frac{\q^2}{q^2}\right)^2\widetilde D_\omega
  \dfrac{1}{1-\Pi_0^V\widetilde D_\sigma\Pi_0^V
    \widetilde D_\omega}
  \\
  \label{eq:Z011}
  D^V&=-\frac{\q^2}{q^2}D_{\sigma\sigma}\Pi_0^V
  \widetilde D_\omega
  \;=\;-\frac{q^2}{\q^2}\widetilde D_\sigma\Pi_0^VD^L\\
  D^T&=\frac{2D_{0\omega}}{1-\frac{1}{2}D_{0\omega}\Pi_0^T}~,
  \intertext{where}
  \widetilde D_\sigma&=\frac{D_{0\sigma}}{1-D_{0\sigma}\Pi_0^S}\\
  \widetilde D_\omega&=\frac{\frac{q^2}{\q^2}D_{0\omega}}{1-
    \frac{q^2}{\q^2}D_{0\omega}\Pi_0^L}~.
  \label{eq:A091}
\end{align}

Observe that the transverse part of the $\omega$-meson is completely decoupled
from the longitudinal propagation, 
that instead involves $\sigma$ and $\omega^0$.

\subsection{The $\sigma$-$\omega$ condensate}
\label{sec:5.2}

We consider first  a simplified model where only the $\sigma$ meson
exists. Since the $\sigma$-exchange is attractive, a $\sigma$ condensation
is expected under certain conditions. In this model the RPA $\sigma$ propagator
\eqref{eq:A081} reads
\begin{equation}
  \label{eq:Z004}
  D_{\sigma\sigma}=\frac{D_{0\sigma}}{1-D_{0\sigma}\Pi_0^S}
\end{equation}
and a $\sigma$ condensate arises if the denominator of the above equation
vanishes at a certain $\q$ and at $q_0=0$, i.e., if the equation
\begin{equation}
  \label{eq:Z005}
  1-D_{0\sigma}(\q,q_0=0)\Pi_0^S(\q,q_0=0)=0
\end{equation}
admits some solution in $\q$.

Let us now study the function $D_{0\sigma}(\q,q_0=0)\Pi_0^S(\q,q_0=0)$.
From eq.~\eqref{eq:067} we get
$$D_{0\sigma}(\q,q_0=0)=-\frac{g_\sigma^2}{\q^2+m_\sigma^2}~,$$
while an easy calculation, using the explicit expressions given in 
\cite{BaCeQu-05}, provides
\begin{multline}
  \label{eq:Z006}
  \Pi_0^S(\q,q_0=0)=-\frac{4m^2+\q^2}{2\pi^2\q}\Biggl\{
 \q\log\frac{k_F+E_F}{m}\\+E_F
  \log\left|\frac{2k_F+\q}{2k_F-\q}\right|
  -\frac{1}{2}\sqrt{4m^2+\q^2}
  \log\left|\frac{E_F\q+k_F\sqrt{4m^2+\q^2}}{E_F\q-k_F\sqrt{4m^2+\q^2}}
    \right|\Biggr\}~.
\end{multline}
A simple check shows that this function is regular at $\q=2k_F$, 
and from the above it immediately follows that
\begin{multline}
  \label{eq:Z007}
  \lim_{\q\to0}D_{0\sigma}(\q,q_0=0)\Pi_0^S(\q,q_0=0)
  =\frac{g_\sigma^2}{m_\sigma^2}\frac{2m^2}{\pi^2}\log\frac{k_F+E_F}{m}
  >0~.
\end{multline}
On the other hand  for large $\q$'s $\Pi_0^S(\q,q_0=0)$ has 
a finite limit so that 
\begin{equation}
  \label{eq:Z008}
  D_{0\sigma}(\q,q_0=0)\Pi_0^S(\q,q_0=0)\underset{|{\bf q}|
    \to\infty}\longrightarrow \q^{-2}~.
\end{equation}
As a consequence eq. \eqref{eq:Z005} has certainly a solution
provided
\begin{equation}
  \label{eq:Z009}
\frac{g_\sigma^2}{m_\sigma^2}\frac{2m^2}{\pi^2}\log\frac{k_F+E_F}{m}>1~.
\end{equation}

The physics contained in this conclusion was  expected: a sufficiently
large attraction produces a $\sigma$ condensate. Numerically
at the normal nuclear density ($k_F=1.36 {\rm fm}^{-1}$) and with
$m_\sigma=550~{\rm MeV}$ used in QHD~\cite{SeWa-86} we get for $g_\sigma$ the critical value
$g_\sigma=2.47$, well below the value of the QHD, namely $g_\sigma=9.573$.

Actually the $\omega$-meson exchange, being repulsive, could prevent
such an occurrence. A simple algebra shows
that $D_{\sigma\sigma}$, $D^L$ and $D^V$, as given in eqs. \eqref{eq:A081},
\eqref{eq:Z010} and \eqref{eq:Z011}, have the same denominator, namely
\begin{equation}
  {\frak D}_L=\left[1-D_{0\sigma}\Pi^S\right]\left[1-(\zeta-2)D_{0\omega}\Pi^L
      \right]
    -(\zeta-2)D_{0\sigma}
    D_{0\omega}\left(\Pi^V\right)^2
    \label{eq:A099}
\end{equation} 
(the index $L$ reminds us that we are concerned with the longitudinal
propagation),
where we have introduced the shortcut
\begin{equation}
  \label{eq:A087}
  \zeta=\frac{\q^2+q_0^2}{\q^2}\qquad\Longrightarrow\qquad 
\zeta-2=\frac{q^2}{\q^2}~.
\end{equation}
The occurrence of a phase transition is now signalled by the existence
of solutions of the equation
\begin{equation}
  \label{eq:Z012}
  {\frak D}_L=0~.
\end{equation}
Proceeding as before we find
\begin{align}
  \label{eq:Z013}
  \Pi^V(\q,q_0=0)&=-\frac{m}{(2\pi)^2\q}\left\{4k_F\q+(4k_F^2-\q^2)\log\left|
      \frac{2k_F+\q}{2k_F-\q}\right|\right\}\\
  \Pi^L(\q,q_0=0)&=-\frac{1}{6\pi^2\q}\Biggl\{2\q k_F E_F+
 2\q(3m^2-\q^2)\log\frac{k_F+E_F}{m}\\
 \nonumber
 &+E_F(4E_F^2-3\q^2)
  \log\left|\frac{2k_F+\q}{2k_F-\q}\right|\\
  \nonumber
  &-\sqrt{4m^2+\q^2}(2m^2-\q^2)
  \log\left|\frac{E_F\q+k_F\sqrt{4m^2+\q^2}}{E_F\q-k_F\sqrt{4m^2+\q^2}}
    \right|\Biggr\}
\end{align}
together with the relevant limits for $\q\to0$
\begin{align}
  \label{eq:Z015}
  \lim_{\q\to0}\Pi^V(\q,q_0=0)&=-\frac{2mk_F}{\pi^2}\\
  \label{eq:Y015}
  \lim_{\q\to0}\Pi^L(\q,q_0=0)&=
  -\frac{1}{\pi^2}\left\{k_FE_F+m^2\log\frac{k_F+E_F}{m}
  \right\}~,
\end{align}
both limits being negative,
and for $\q\to\infty$
\begin{align}
  \label{eq:Z017}
  \Pi^V(\q,q_0=0)&\sim\frac{1}{\q^2}\\
  \label{eq:Y017}
  \Pi^L(\q,q_0=0)&\to {\rm constant}~.
\end{align}
Moreover in the limit $q_0\to0$ we find
$$\zeta=1~,\qquad\qquad\zeta-2=-1~.$$
 Thus at $q_0=0$
\begin{equation}
  {\frak D}_L\bigm|_{q_0=0}=\left[1-D_{0\sigma}\Pi^S\right]\left[1+
  D_{0\omega}\Pi^L
      \right]
    +D_{0\sigma}
    D_{0\omega}\left(\Pi^V\right)^2~.
    \label{eq:Z018}
\end{equation} 
which implies, using \eqref{eq:Z067}, \eqref{eq:Z008}, \eqref{eq:Z017}
and \eqref{eq:Y017}
\begin{equation}
  \label{eq:Z019}
  \lim_{\q\to\infty}{\frak D}_L(\q,q_0=0)=1~.
\end{equation}
Thus a condensed state will occur if 
\begin{equation}
  \label{eq:Z020}
  \lim_{\q\to0}{\frak D}_L(\q,q_0=0)\leq0~.
\end{equation}
Let now consider the above limit as a function of $k_F$:
\begin{equation}
  \label{eq:Z021}
  \phi(k_F)=\lim_{\q\to0}{\frak D}_L(\q,q_0=0)~.
\end{equation}
From \eqref{eq:Z008},\eqref{eq:Z015}
and \eqref{eq:Y015} we  derive, for low $k_F$,
\begin{equation} 
  \label{eq:Z022}
  \phi(k_F)=1-\frac{2m}{\pi^2}\left(\frac{g_\sigma^2}{m_\sigma^2}-
    \frac{g_\omega^2}{m_\omega^2}\right)k_F+{\cal O}(k_F^2)~,
\end{equation}
while
at large $k_F$ we find the asymptotic behaviour
\begin{equation}
  \label{eq:Z023}
  \phi(k_F)\simeq-\frac{2}{\pi^4}
  \frac{g^2_\sigma}{m_\sigma^2}\frac{g_\omega^2}{m_\omega^2}
  m^2k_F^2\log\frac{2k_F}{m}+\frac{1}{\pi^4}\frac{g_\omega^2(4g^2_\sigma
    m^2+\pi^2m_\sigma^2)}{m_\sigma^2m_\omega^2}k_F^2~.
\end{equation}
Hence $\phi(k_F)$ goes to 1 for low $k_F$ and to $-\infty$ for large $k_F$
and thus a critical value for $k_F$ surely exists. 
The problem of how and when this critical point is reached and this can 
be addressed only numerically.

\begin{figure}[ht]
  \begin{center}
    \leavevmode
    \epsfig{file=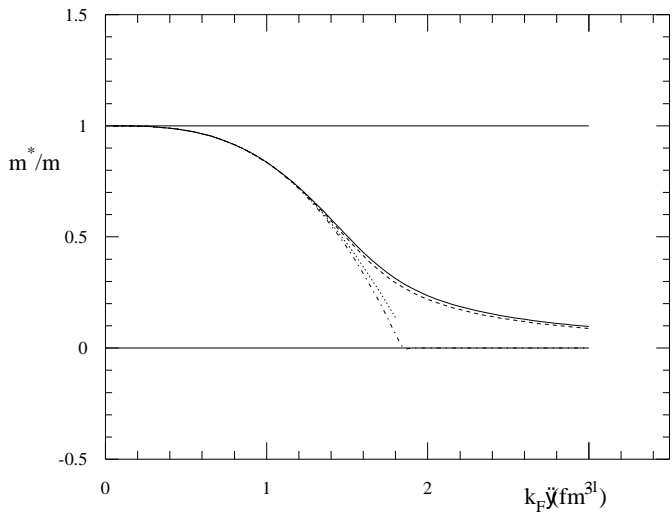,width=12cm,height=8cm}
    \caption{Nucleon effective mass. Solid line: full self-consistent 
      calculation; dashed line: self-consistent calculation with $a_4=0$; 
      dotted line: non- self-consistent calculation, dash-dotted line: 
      non- self-consistent calculation with $a_4=0$.}
    \label{fig:6}
  \end{center}
\end{figure}

Before exploiting the calculations, we observe that
up to now we have not accounted for self-consistency in determining
the nucleon mass, according to eq. \eqref{eq:Y001}. 
The nucleon effective mass $m^*$ is displayed in fig. \ref{fig:6}
and compared with a non-self-consistent calculation (i.e., with
$S_H$ replaced by $S_0$ in eq.~\eqref{eq:064}). The plot shows,
as pointed out by Lee and Wick many years 
ago\cite{LeWi-74}, that the self-consistent effective mass tends to vanish at
large densities. 
Thus the large $k_F$ limit \eqref{eq:Z023}
must be  evaluated at $m=0$, namely
\begin{equation}
  \label{eq:Z024}
  \phi(k_F)\simeq\frac{g_\omega^2}{m_\omega^2}\frac{k_F^2}{\pi^2}>0~.
\end{equation}

\begin{figure}[ht]
  \begin{center}
    \leavevmode
    \epsfig{file=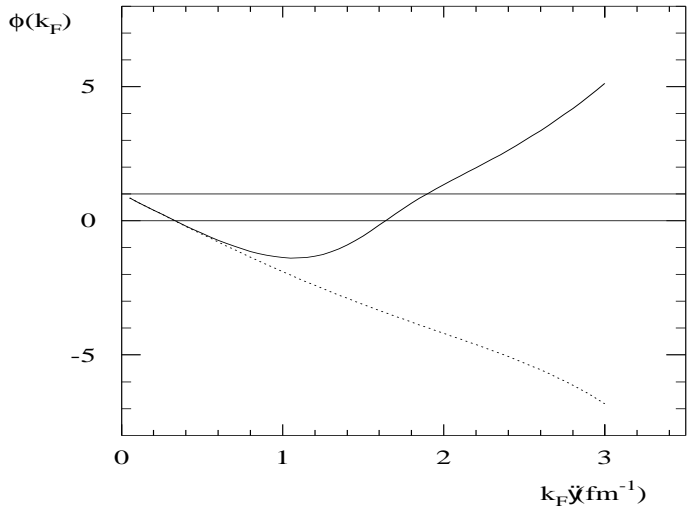,width=12cm,height=8cm}
    \caption{Plot of the function $\phi(k_F)$ with a self-consistent 
      effective mass (solid line) and the bare mass (dashed line).
      The last is evaluated with the standard values of QHD.}
    \label{fig:1cond}
  \end{center}
\end{figure}

On the other hand, at low $k_F$ the effective mass tends to the bare one and
with the standard parameters of QHD, namely $g_\sigma=9.573$, $m_\sigma=550$ 
MeV, $g_\omega=11.67$ and $m_\omega=783$ MeV,
 $\phi(k_F)$ start decreasing if
$$\frac{g_\sigma^2}{m_\sigma^2}-\frac{g_\omega^2}{m_\omega^2}>0~.$$
Thus one may reasonably argue that the function may have a minimum 
below 0. An explicit calculation confirms this guess, as shown in fig. 
\ref{fig:1cond}.

Actually, to be reasonably sure that no condensation arises, we could require
$$\phi'(k_F)\bigm|_{k_F=0}\geq0$$
that corresponds to a limiting value $g_\sigma=8.197$, to be compared with
$g_\sigma=9.573$ as given in \cite{SeWa-86}.

\begin{figure}[ht]
  \begin{center}
    \leavevmode
      \epsfig{file=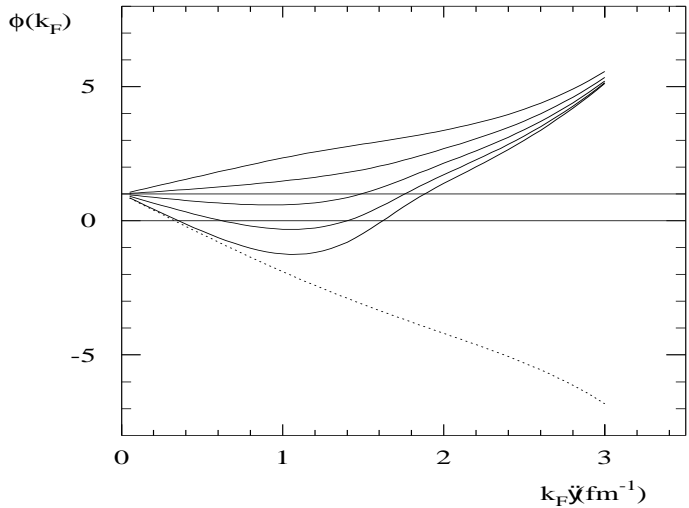,width=12cm,height=8cm}  
    \caption{The function  $\phi(k_F)$ for different values of $g_\sigma$, 
      ranging from 10 (lowest solid line) to 8 (upper one) with a 
      spacing of 0.5, with self-consistent effective mass. The dashed line
      corresponds to a bare nucleon mass.}
    \label{fig:2cond}
  \end{center}
\end{figure}

A more detailed search for limiting conditions requires a numerical 
calculation. In fig. \ref{fig:2cond} we show the behaviour of $\phi$
for five different values of $g_\sigma$ ranging from $g_\sigma=10$ (lowest
solid line) to $g_\sigma=8$ (upper solid line).
It turns out that the critical value
for the existence of a $\sigma$- condensate is
\begin{equation}
  \label{eq:X001}
  {g_\sigma}_{\rm crit}=8.828~.
\end{equation}
It is clear nevertheless that close to the critical 
value precursor phenomena (like in pion condensation) are to be expected,
thus a lower value for $g_\sigma$ should be preferred.

\begin{figure}[ht]
  \begin{center}
    \leavevmode
      \epsfig{file=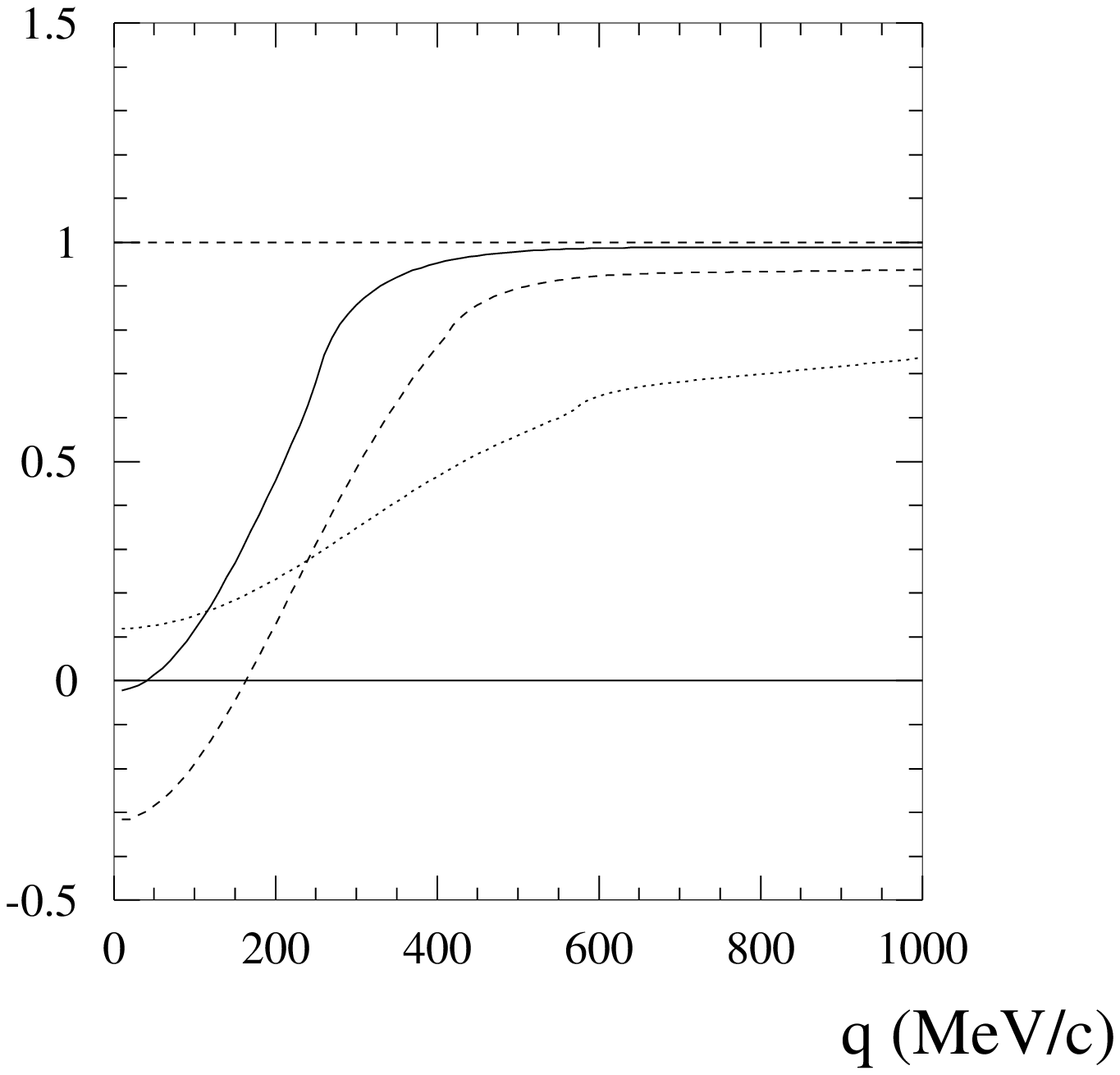,width=12cm,height=8cm}  
    \caption{The function  ${\frak D}_L\vert_{q_0=0}$ for different values 
      of $k_F$.
      Solid line: $k_F=127.8$ MeV/c, dashed line: $k_F=207.8$ MeV/c,
      dotted line $k_F=287.8$ MeV/c.}
    \label{fig:4}
  \end{center}
\end{figure}

The above states that at $g_\sigma=8.828$ there is (as the curves in fig. 
\ref{fig:1cond} suggest) only one value of $k_F$ at which the 
$\sigma$-condensation occurs. Numerically this happens at 
$k_F^{\rm crit}=207.8$ MeV/c. Above this value we see that two solutions
of the equation $\phi(k_F)=0$ exist. Thus below the lowest solution
we find no $\sigma$-condensation, and the same occurs above the higher one.
In fig. \ref{fig:4} we have plotted the function ${\frak D}_L\vert_{q_0=0}$
for three different values of $k_F$.
This shows how below the lower value and above the higher one no
phase transition seems to occur.
Actually this statement is only formal: indeed when a phase
transition is found at the lower critical value of $k_F$, then what happens
for higher $k_F$ becomes unpredicable in the present formalism.
On the other hand since no evidence exists about this phase transition, 
the existence of another critical value of $k_F$ is simply of mathematical,
but not physical, interest.

\section{Conclusions}
\label{sec:9}

The QHD was taylored to describe the static properties of the nuclear matter
at the mean field level, i.e., at the $0^{\rm th}$ order 
in the loop expansion. 
Within this approximation no phase transition can arise.

We can go beyond, however, and consider an expansion in bosonic loops only
(boson loop expansion, in short BLE), as described in 
\cite{AlCeMoSa-87,CeCoSa-97}. There it was proved that the RPA series is just
the mean field of the BLE: it contains in fact fermionic loops but 
not bosonic ones and in principle it may originate phase transitions,
as is well known from the old discussions about pion condensation.

In this paper we have shown in fact that at 
BLE-mean field level the Serot and Walecka model would
predict the occurrence of a $\sigma$-$\omega$ condensed phase.
This outcome however does not destroy the whole apparatus of QHD:
it simply states that the mean field level is not able to provide
a stable ground state at least above some critical values of $k_F$.
It is remarkable, nevertheless, that these critical values lie below 
the normal nuclear density.

To overcome this instability we need to perform a higher order calculation
of the binding energy (at the order of two bosonic loops). Since 
the value of the parameters in this model have been fixed at the
mean field level, in going beyond we need a reparametrization, that
hopefully  should lead to a non-critical choice.

To conclude we observe that in the present model we  have studied
the possible occurrence of a condensed state as $g_\sigma$ varies.
We could as well modify the $\sigma$ mass, as the relevant parameter is 
their ratio. Actually this is ininfluent as far as we look at
the existence of a critical point, however, it matters when we consider 
the momentum ${\bf q}$ where the condensation occurs.

\bibliography{references}

\end{document}